\documentclass[11pt, a4paper]{article}
\usepackage[english]{babel}
\usepackage[latin1]{inputenc}
\usepackage{epsfig}

\begin{document}
\sloppy
\title{Thermal dileptons from quark and hadron phases of an expanding fireball}
\author{R. A. Schneider\footnote{e-mail: {\tt schneidr@ph.tum.de}}, W. Weise \\ Physik-Department, Technische Universit\"at M\"unchen, \\
D-85747 Garching, Germany \footnote{Work supported in part by BMBF and GSI} }
\maketitle

\begin{abstract}
A fireball model with time evolution based on transport calculations is used to examine the
dilepton emission rate of an ultra-relativistic heavy-ion collision. A transition from
hadronic matter to a quark-gluon plasma at a critical temperature $T_C$ between 130-170 MeV is assumed. We also consider 
a possible mixed phase scenario. We include thermal corrections to the hadronic spectra 
below $T_C$ and use perturbation theory above $T_C$. The sensitivity of the spectra with respect 
to the freeze-out temperature, the initial
fireball temperature and the critical temperature is investigated.
\end{abstract}
\section{Introduction}
\label{intro}
There are convincing arguments that the theory of the strong interaction, QCD, exhibits a
phase transition at sufficiently high temperatures from a confined hadronic phase to the
quark-gluon plasma (QGP) phase \cite{lattice1}\cite{boyd}\cite{lattice2}. Lattice QCD results set the critical temperature $T_C$ to about 150-200 MeV
\cite{recent1}\cite{recent2}\cite{karsch99}. It is hoped that it is possible to create the QGP in
ultrarelativistic heavy-ion collisions at CERN and BNL, and there are preliminary indications that
the QGP might have already been encountered \cite{ceres2}\cite{ceres3}\cite{qm99}.\\[0.3cm] Dileptons ($e^+e^-$ or
$\mu^+\mu^-$ pairs) are considered to be an optimal probe for the early stages of the collision
because they leave the hot region without thermalization. Hadrons which reach the detector can only
tell us about the later stage of the collision, the freeze-out zone. When the fireball created in a
heavy-ion collision is in the QGP or partonic phase, dileptons are mainly produced by thermal
quark-antiquark annihilation. As the system expands, it cools off and undergoes the
transition into the hadronic phase. There, dileptons come mainly from pion and kaon annihilation
processes, which are dynamically enhanced through the formation of light vector meson resonances
($\rho, \omega$ and $\phi$ with masses below 1.1 GeV). The dilepton emission rate is proportional
to the imaginary part of the hadronic current-current correlation function, or the electromagnetic
spectral function. The invariant mass distribution of the lepton pairs reflects the mass
distributions of the vector mesons at the instant of decay. This offers the possibility to study
the influence of finite temperature and baryonic density on the meson spectral distributions.
\\[0.3cm] In this paper, we calculate the purely thermal dilepton emission rate for
both phases and compare it with data from the CERES/NA45 experiment. Much work has been done to
interpret the observed dilepton enhancement in the low-mass region \cite{ceres1}\cite{ceres2} in terms of
in-medium modifications of hadron spectra \cite{rapp2}\cite{wweise1}\cite{rapp}. We investigate
the thermal modifications that the hadrons may exhibit in the limit of zero baryon density. We demonstrate that these alone can
also explain the data very easily if the initial fireball temperature is sufficiently high to
support a QGP during at least part of the expansion. Finally, we study the dependence of the shape
of the dilepton emission spectra on the parameters of our simple fireball model and discuss a
recently proposed mixed phase structure.
\section{Dileptons from a fireball}
\label{sec:1}
The dilepton emission rate from a hot domain populated by particles in thermal equilibrium at
temperature $T$ is proportional to the imaginary part of the spin-averaged, time-like photon
self-energy, with these particles as intermediate states. The thermally excited particles
annihilate to yield a time-like virtual photon with four-momentum $q$ which decays subsequently
into a lepton-antilepton pair. The differential rate is given by
\begin{equation}
\frac{dN}{d^4 x d^4q}  =  \frac{\alpha^2}{\pi^3 q^2} \ \frac{1}{e^{\beta q^0} - 1} \
\mbox{Im}\bar{\Pi}(q, T),\label{dileptonrate}
\end{equation}
where $\alpha = e^2/4\pi$, $\beta = 1/T$, and we have neglected the lepton masses. We defined
$\bar{\Pi}(q) = -\Pi^\mu_{\ \mu} /3$. Here $\Pi^\mu_{\ \mu}$ denotes the trace over the thermal
photon self-energy which is equivalent to the thermal current-current correlation function
\begin{equation}
\Pi_{\mu\nu}(q,T) = i \int d^4 x \ e^{iqx} \langle \mathcal{T} j_{\mu}(x) j_{\nu}(0) \rangle_\beta,
\end{equation}
where $j_\mu$ is the electromagnetic current. The result eq.(\ref{dileptonrate}) is valid to order $\alpha$ in the electromagnetic interaction
and to all orders in the strong interaction.
\\[0.3cm] To compare with experimental data, we set up a model for the space-time evolution of a heavy-ion
collision, assuming approximate thermal equilibrium to be a useful concept \cite{equ1}. Some recent
discussions suggest \cite{equ2}\cite{equ3} that equilibration times at the conditions of heavy ion
collisions at CERN may indeed be very short (less than a fermi), small compared to expansion
times of order 10 fm/$c$, although this is still a matter of debate. The easiest approach for our purposes is to use a simplified fireball
model which parametrizes the time dependence of temperature and volume in accord with microscopic
transport calculations \cite{li1}\cite{li2}. This assumes that the fireball can be characterized by a
homogeneous temperature at all times and that it cools off adiabatically with the parametrization
\begin{equation}
T(t) = (T^i - T^\infty) e^{-t/\tau_1} + T^\infty. \label{Tt}
\end{equation}
This ansatz introduces the initial temperature $T^i$, a time constant $\tau_1$ and an asymptotic
temperature $T^\infty$. Furthermore, we assume an isotropic expansion of the fireball in the
centre-of-mass frame of the collision such that the time evolution of the volume can be described
by\begin{equation}
V(t) = \frac{N_B}{\rho(t)}, \qquad \rho(t) = \rho^i e^{-t/\tau_2}, \label{rhot}
\end{equation}
where $N_B$ is the number of baryons which participate in the reaction, $\tau_2$ is another time
constant and $\rho^i$ is the initial baryon density of the hot spot. More sophisticated volume
parametrizations can be used, but they introduce more parameters and the differences in the resulting
integrated rates are marginal.
\\[0.3cm]
Finally, to obtain the measured dilepton rates, we integrate eq.(\ref{dileptonrate}) over the
space-time history of the collision to compare them with the CERES/NA45 data.  The CERES experiment is a {\em fixed-target}
experiment, i.e. the dilepton rates are measured in the lab frame whereas our volume
parametrization (\ref{rhot}) is valid for a fireball at rest, in the c.m. frame of the collision.
In the lab frame, the CERES detector covers a limited rapidity interval $\eta = 2.1-2.65$. The
rapidity needed in order to boost to the c.m. frame is therefore 2.375. The CERES collaboration chose to
display its data in the format $d^2N/dMd\eta$, where $\eta$ is the (pseudo-)rapidity of the virtual
photon. The longitudinal momentum $p_L$ is related to $\eta$ by $p_L = m_T \sinh \eta$ with the
'transverse mass' $m_T = \sqrt{M^2 + p_T^2}$. To be able to compare our calculated rates to the data, we integrate these rates over the transverse momentum $p_T$ only, given that $$d^4p = M p_T \ dM
 \ d\eta \ dp_T \ d\theta.$$  This is equivalent to integrating over all three-momenta as long as we set $\eta$ = 0 in the c.m. frame. 
\\[0.3cm]
The formula for the space-time- and $p_T$-integrated dilepton rates is now
\begin{eqnarray}
\frac{d^2N}{dM d\eta} = & & 2\pi M \int \limits_0^{t_{f}} dt \frac{N_B}{\rho(t)} \ \int
\limits_0^\infty dp_T \ p_T \nonumber \\ & &  \cdot \frac{dN(T(t),M, \eta, p_T)}{d^4 x d^4p} \ Acc(M, \eta, p_T),
\label{integratedrates}
\end{eqnarray}
where $t_{f}$ is the freeze-out time of the collision, and the matrix $Acc(M, \eta, p_T)$ accounts for
the experimental acceptance cuts specific to the detector. At the CERES experiment, each
electron/positron track is required to have a transverse momentum $p_T > 0.2$ GeV, to fall into the
rapidity interval $2.1 < \eta < 2.65$ in the lab frame and to have a pair opening angle
$\Theta_{ee} > 35$ mrad. Finally, for comparison with the CERES data, eq.(\ref{integratedrates})
has to be divided by $dN/d\eta$, the number of particles per unit rapidity. As the data have been
taken close to midrapidity, i.e. around the central plateau of the approximate Gaussian
distribution of $dN/d\eta$ in the c.m. frame as mentioned above, its value is assumed to be
constant over the $\eta$ range restricted by the detector cuts.
\section{Spectrum above critical temperature}
\label{sec:2}
Lattice QCD calculations suggest \cite{lattice1}\cite{lattice2} that sufficiently far above the critical
temperature $T_C$ for the deconfinement transition, quarks and gluons can be treated approximately
as an ideal gas; they interact only weakly because of the behaviour of the running coupling
strength $\alpha_s$ at high $T$. In this temperature region we can evaluate the hadronic part of
the photon self-energy using perturbative QCD. To lowest order in $\alpha$ and to zeroth order in
$\alpha_s$, $\Pi_{\mu\nu}$ corresponds simply to the thermal quark-antiquark loop to which we restrict ourselves in the present paper.  Higher order thermal QCD corrections beyond the leading $q \bar{q}$ loop will be investigated in forthcoming work.
\\[0.3cm] At our typical temperatures $T$ of several hundred MeV, we need to take into account
the three flavours
$u$, $d$ and $s$. Heavier quarks such as the $c$ and $b$ are strongly suppressed by Boltzmann
factors because of their large mass and can be considered as 'frozen'. The total contribution to
the dilepton emission rate is obtained by summing the individual $q\bar{q}$ contributions
multiplied by the squares of the electric charges $e_f$ of their respective flavour currents. In
addition, each quark exists in three colours which leads to an overall multiplicative factor of
$N_C = 3$. Furthermore, the $u$ and $d$ quarks are considered massless and the heavier
$s\bar{s}$-pair can only be produced if $q^2$ is larger than $4m_s^2$. So the total imaginary part
of the self-energy which will be used in eq.(\ref{dileptonrate}), evaluated with standard
finite-temperature techniques \cite{LeBellac}, is
\begin{eqnarray}
\mbox{Im}\bar{\Pi}(q,T) & = & \frac{q^2}{4\pi} \sum_{f=u,d,s} e_f^2 \ \theta(q^2 - 4m_f^2) \left( 1
+ \frac{2m_f^2}{q^2}\right) \sqrt{1 - \frac{4m_f^2}{q^2}} \nonumber \\ & & \cdot \  
 \left[ \frac{2T}{|{\bf q}|} \ \left(1 - \frac{4m_f^2}{q^2}\right)^{-\frac{1}{2}}
\ln \frac{f_D(E_-)}{f_D(E_+)} -1 \right], \label{ImPiqqbar}
\end{eqnarray}
where $$f_D(E) = \frac{1}{\exp(\beta E) +1}$$ and $$E_\pm = \frac{q^0}{2} \pm \frac{| \vec{q}| }{2}\sqrt{1
- \frac{4m_f^2}{q^2}}.$$In the limit $T \rightarrow 0$, eq.(\ref{ImPiqqbar}) reduces to the
well-known vacuum result. At finite $T$, the quantity (\ref{ImPiqqbar}) is always smaller than
Im$\bar{\Pi}(q, T=0)$ as a consequence of Pauli blocking.
\section{Spectrum below critical temperature}
\label{sec:3}
At $T < 150$ MeV, confinement sets in and the effective degrees of freedom of the QCD Hamiltonian
are now colourless hadrons. The
photon can directly couple to $J^P = 1^-$ states (the lowest ''dipole'' excitations of the
QCD vacuum). In the energy region of
interest, these particles are the $\rho$, $\omega$ and $\phi$ mesons and multi-pion states carrying the same quantum numbers. Our next step is thus to
connect the electromagnetic current-current correlation function with the currents generated by these
mesons. We use an effective Lagrangian which approximates the $SU(3)$ flavour sector of QCD at low
energies. The appropriate model for our purposes is the {\em improved Vector Meson Dominance} model
combined with chiral dynamics of pions and kaons as described in \cite{wweise1}.\\[0.3cm] Within
this model, the following relation between the imaginary part of the irreducible photon self-energy
$\mbox{Im} \bar{\Pi}$ and the vector meson self-energies $\Pi_V(q)$ in vacuum can be derived:
\begin{eqnarray}
\mbox{Im} \bar{\Pi}(q) & = & \sum \limits_V \frac{\mbox{Im} \Pi_V(q)}{g_V^2} \ |F_V(q)|^2,
\label{ImBarPi}
\\ F_V(q) & = & \frac{\left( 1- \frac{g}{g^0_{V}} \right)q^2 - m_V^2}{q^2 - m_V^2
+ i \mbox{Im}\Pi_V(p)},
\end{eqnarray}
where $m_V$ are the (renormalized) vector meson masses \footnote{In the present context, $m_V$ is understood to include the shift from bare to physical mass induced by Re$\Pi_V$ \cite{wweise1}.}, $g^0_V$ is the $\gamma V$ coupling and $g$ the $\Phi V$
coupling, where $\Phi$ stands for one of the pseudoscalar Goldstone bosons $\pi^\pm, \pi^0$ and
$K^\pm, K^0$. Eq.(\ref{ImBarPi}) is valid for a virtual photon with vanishing three-momentum,
$q = (q^0, 0,0,0)$. Taking the limit $|\vec{q}| = 0$ should be reasonable for our purposes in view of the fact that the c.m. rapidity interval accessible at CERES restricts $|\vec{q}|$ on average to a fraction of the vector meson mass $m_V$. For finite three-momenta there would be two scalar functions $\bar{\Pi}_L$ and
$\bar{\Pi}_T$, because the existence of a preferred frame of reference (the heat bath) breaks
Lorentz invariance. In the following, we calculate the hadronic spectra $\Pi_V$ at finite
temperature.
\subsection{The $\rho$ meson spectrum}
The main decay channel of the $\rho$-meson is $\rho \rightarrow \pi^+\pi^- $ with a vacuum width
$\Gamma \simeq$ 150 MeV. In the VMD model, hadronic current conservation
leads to two possibilities to couple the $\rho$ to pions: There is a pion tadpole diagram and a two
pion loop diagram. Both terms contribute to $\Pi_{\rho}$ and correspondingly modify the $\rho$
meson properties. Their vacuum values are well-known \cite{wweise1}, thus in the following we will
calculate only the changes induced by finite temperature.\\[0.3cm] Using the real-time formalism of
thermal field theory \cite{LeBellac}, we find for the imaginary part of the two pion loop diagram
\begin{eqnarray}
& & \mbox{Im} \bar{\Pi}(q^0, \vec{q}=0) =  - (q^0)^2 \frac{g_{\rho\pi\pi}^2}{48\pi} \left( 1 -
\frac{4m_{\pi}^2}{(q^0)^2} \right) ^{\frac{3}{2}} \left(1 + 2 f_B(q^0/2) \right) \theta(q^0 -
2m_\pi). \label{ImRhoPiPi}
\end{eqnarray}
This is exactly the $T=0$ result multiplied by the characteristic thermal factors: the imaginary part
is larger than at zero temperature as a consequence of the Bose-enhancement of the thermal pions. Using the
relation \mbox{$\mbox{Im}\bar{\Pi}(m_\rho) = m_\rho \Gamma$}, we see that the decay width $\Gamma_{\rho
\rightarrow \pi\pi}$ of the $\rho$ meson is enhanced because there are already thermally excited
pions present in the heat bath. At temperatures of about 100 MeV, this effect causes an increase of
the $\rho$ width by about 20\% . Additional sources of thermal broadening at high temperatures are reactions of the $\rho$ meson with pions in the heat bath, such as $\rho \pi \rightarrow 3\pi$. We checked that this is only a small correction to the width.
Other decay modes of the
$\rho$ like $\rho \rightarrow \pi^+ \pi^- \gamma$, not taken into account, may also start to play a role at high $T$, however, their
branching ratios are very small.
\\[0.3cm] The calculation of the real part of the thermal $\rho$
meson self-energy is technically more involved, but the resulting temperature dependent mass shift
$\delta m_\rho(T)$ is small. Eletsky and Ioffe \cite{ioffe1}\cite{ioffe2} have pointed out
that the leading term of $\delta m_\rho$ starts at order $T^4$ when $\rho - a_1$ mixing is properly
taken into account. The actual evaluation of $\delta m_\rho$ gives an (upward) mass shift of less
than 3\% of $m_\rho$ even at temperatures as high as 150 MeV. We can therefore safely neglect this
effect in our calculation.
\\[0.3cm] Note that the general form of the thermal $\rho$ spectrum does not change dramatically, at least up to $T \leq$ 150 MeV. Substantial broadening of the $\rho$ meson spectrum is expected in the presence of large baryon density \cite{wweise1}\cite{rapp}. However, under the conditions of Pb-Au collisions at CERN, the fireball rapidly expands and the baryon density at hadronization drops to a fraction of normal nuclear matter density. We can thus ignore such density effects. 

\subsection{The $\omega$ meson spectrum}
The main decay mode of the $\omega$(783) vector meson with a branching ratio of 89\% is $\omega
\rightarrow \pi^+ \pi^0 \pi^-$ and $\Gamma_{\omega \rightarrow 3\pi}$ = 7.5 MeV. It is well-known
that a point-like coupling $\omega \rightarrow 3\pi$ does not describe the $\omega$ decay properly.
It is necessary to include an intermediate step, the Gell-Mann, Sharp, Wagner (GSW) process
\cite{GSW}  $\omega \rightarrow \rho \pi \rightarrow \pi^+ \pi^- \pi^0.$ The full decay process is
described by an interference between the amplitudes of these two processes. In the following, we
assume that the intermediate virtual rho meson propagation is merely a 'vertex' correction and that
the effects of temperature on the $\rho$ meson are small, so we can neglect them to leading order.
This is justified by noting that the thermal production of a $\rho$ meson is heavily suppressed by
$e^{-m_\rho/T} \ll 1$ in the temperature range we are interested in (up to about 150 MeV).
\\[0.3cm] The explicit calculation \cite{ras1} for an $\omega$ meson at rest,
$q_\omega = (\omega, \vec{q} = 0)$, leads us to the result
\begin{eqnarray}
\mbox{Im}\bar{\Pi}(\omega) & = &  \frac{1}{192 \pi^3} \left( \mathcal{B}_1 + 3 \ \mathcal{B}_2
\right) , \label{gsw_omega}
\\ \mathcal{B}_1 & = & \int \limits_{\Delta_1} dE_+ dE_- \left( \vec{q}_+^{\ 2} \vec{q}_-^{\ 2} -
(\vec{q}_+ \cdot \vec{q}_-)^2 \right) \ |\mathcal{M}(\omega; q_+, q_0, q_-)|^2 \cdot n(E_+, E_-,
\omega), \nonumber \\ \mathcal{B}_2 & = & - \int \limits_{\Delta_2} dE_+ dE_- \left( \vec{q}_+^{\ 2}
\vec{q}_-^{\ 2} - (\vec{q}_+ \cdot \vec{q}_-)^2 \right) \ |\mathcal{M}(\omega; q_+, -q_0, q_-)|^2 \cdot
n(E_+, E_-, \omega), \nonumber
\end{eqnarray}
where 
\begin{eqnarray*}
& & \mathcal{M}(\omega; q_+, q_0, q_-)  =   \frac{-3h}{f_\pi^3} + 2 \ \frac{g_{VVP}}{f_\pi} \
g_{\rho \pi \pi} \sum_{\alpha = +, 0, -} \frac{1}{(q_\omega-q_\alpha)^2 - m^2_\rho -
\bar{\Pi}_\rho(q_\omega -q_\alpha)}.
\end{eqnarray*}
 Here, $\bar{\Pi}_\rho$ is the zero-temperature rho meson
self-energy. Its explicit expression and a general discussion of this matrix element can be found
in \cite{wweise1}, including the coupling constants $h$ and $g_{VVP}$ ($f_\pi$ is the pion decay constant). $E_\pm$ and $\vec{q}_\pm$ are the energies and momenta of the charged pions in
the final state. To keep our notation short, we have introduced the function
\begin{eqnarray*}
n(E_+, E_-, \omega)  & = & \Big( (1 + f_B(E_+)) (1 + f_B(E_-))(1 + f_B(\omega - E_+ - E_-))
 - \\ & & - f_B(E_+) \ f_B(E_-) \ f_B(\omega - E_+ - E_-) \Big).
\end{eqnarray*}
In eq.(\ref{gsw_omega}), $\mathcal{B}_1$ describes the process of a Bose-enhanced three-particle
decay of the $\omega$ meson, $n(E_+, E_-, \omega)$ accounts for the characteristic Bose factors.
The kinematically allowed integration region in the $E_+ E_-$ plane, $\Delta_1$, is the same as in
the $T=0$ case for the decay process. It is limited to $m_\pi \le E_\pm \le \omega - 2m_\pi$ and
$2m_\pi \le E_+ + E_- \le \omega - m_\pi$ with the constraint $ \vec{q}_+^{\ 2} \vec{q}_-^{\ 2} -
(\vec{q}_+ \cdot \vec{q}_-)^2 > 0$.\\[0.3cm] To interpret the $\mathcal{B}_2$ term in
eq.(\ref{gsw_omega}), we note that
\begin{eqnarray}
 - n(E_+, E_-, \omega)  & = & \Big( (1 + f_B(E_+)) (1 + f_B(E_-)) \  f_B(E_+ + E_- - \omega) - \nonumber \\ & &
- f_B(E_+) \ f_B(E_-) \ (1 + f_B(E_+ + E_- - \omega)) \Big),\label{minusn}
\end{eqnarray}
by using $1 + f_B(E) + f_B(-E) = 0$. Thus $\mathcal{B}_2$ corresponds to the scattering of
the $\omega$ meson off a thermally excited $\pi^0$ into $\pi^+$ and $\pi^-$, where each particle
has the characteristic thermal corrections. The integration region $\Delta_2$ is still constrained
by $ \vec{q}_+^{\ 2} \vec{q}_-^{\ 2} - (\vec{q}_+ \cdot \vec{q}_-)^2 > 0$, but now in the region defined by
$E_+ + E_- \ge \omega + m_\pi$. Because the integration region $\Delta_2$ is now unbounded and the
functions involved are only damped by Bose factors and their combinations, the $\omega \pi \rightarrow \pi \pi$ scattering
contribution to the decay width increases quite strongly with temperature, as we will see below.
The factor 3 multiplying $\mathcal{B}_2$ comes from summing over all possible charge combinations $\omega +
\pi^+ \rightarrow \pi^0 + \pi^+$, $\omega + \pi^- \rightarrow \pi^0 + \pi^-$ and $\omega + \pi^0 \rightarrow \pi^- + \pi^+$.
\\[0.3cm] The evaluation of the real part of the $\omega$ meson self-energy coupled to the $3\pi$
continuum is much more difficult than that of the imaginary part because it would involve a
two-dimensional Cauchy principal value integral. We refrain from calculating the real part of the
three-body decay, assuming that its temperature dependent mass shift is not much different from the
(small) one of the $\rho$ meson.

\subsection{The $\phi$ meson spectrum}
Because of the large mass of the $\phi$ and its main two-body decays into heavy kaons, the impact
of temperature on its spectrum is marginal. The thermal production of kaons at temperatures even as
high as 150 MeV is still suppressed by an order of magnitude compared to thermal pion production.
We find that the sharp peak structure of the $\phi$ meson does not change as its total width
increases by about 20\% to just 5.2 MeV at the highest temperatures. Thus it is only moderately
affected by temperature effects and one should be able to observe its sharp resonance structure in
future dilepton rate measurements.

\section{Integrated rates}

\begin{figure}[h]
\begin{center}
\epsfig{file=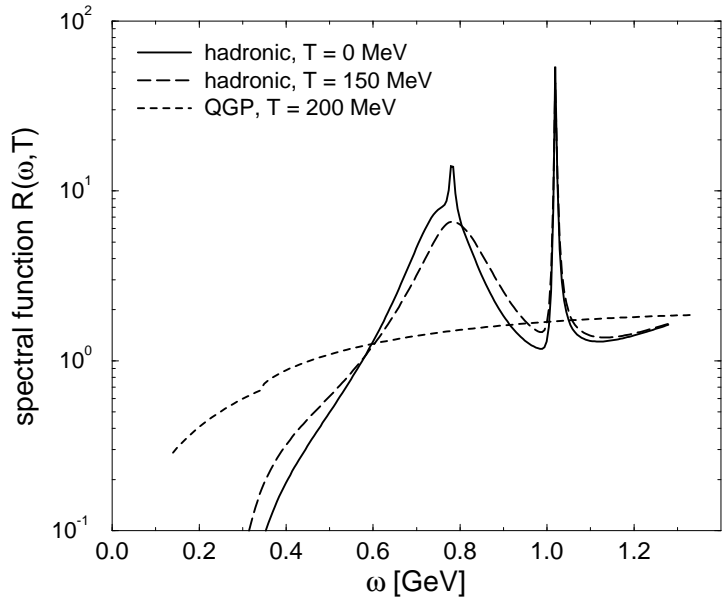,width=9cm, angle=-90} \caption{Spectral distribution $R =
12\pi \mbox{Im}\bar{\Pi}/q^2$ for $q^\mu = (\omega, \vec{q} = 0)$ for different temperatures above and below the quark-hadron transition. The background contribution to the hadronic spectra at large $\omega$ arises from the 4$\pi$ continuum.} \label{all_spectra}
\end{center}
\end{figure}

\begin{figure}
\begin{center}
\epsfig{file=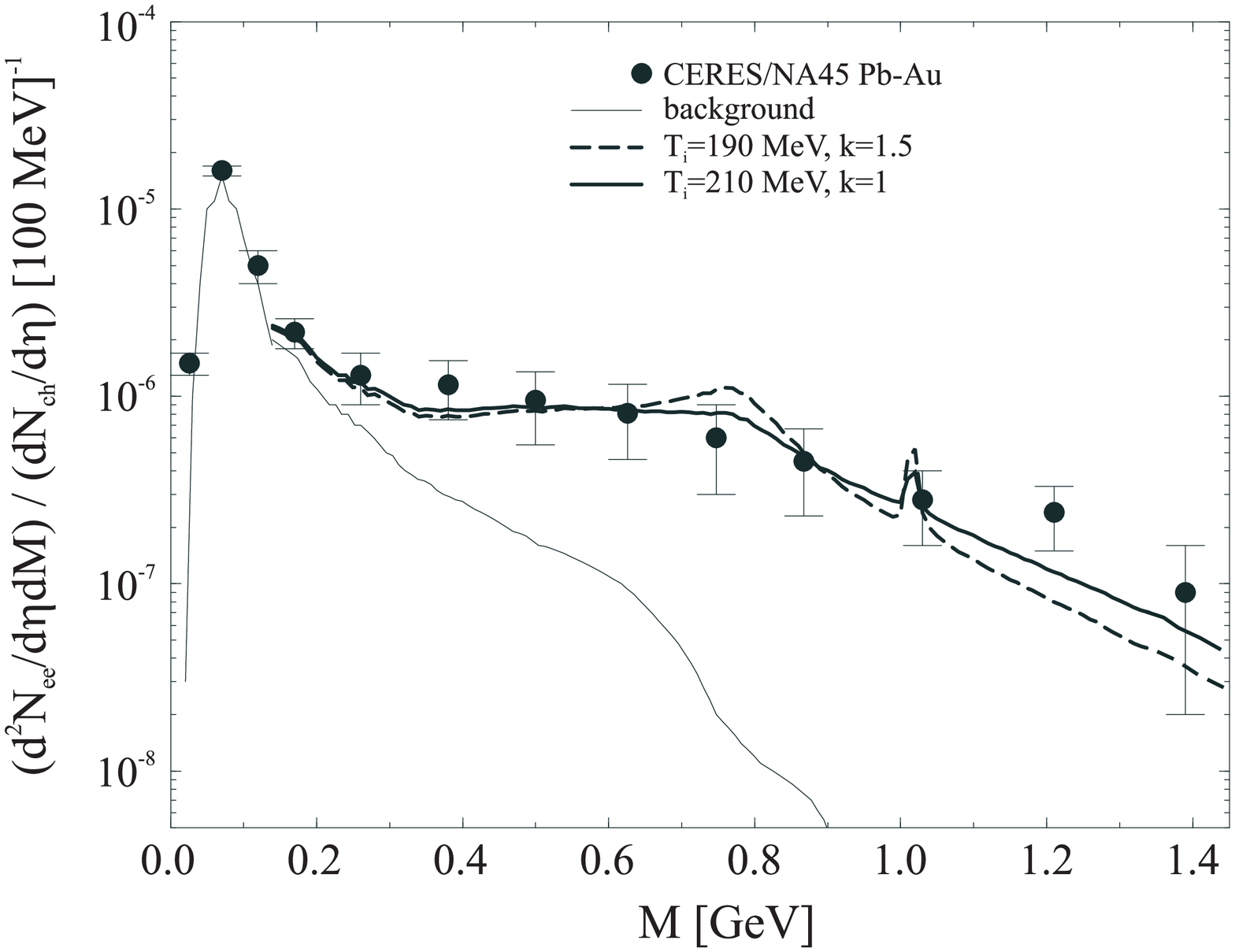,width=10cm} \caption[Dilepton rates as a function of the invariant $e^+e^-$ mass $M$, calculated in a mixed scenario (hadrons below $T_C$, quarks above $T_C$ = 150 MeV) for
the Pb-Au case with two different initial temperatures]{A mixed scenario for the Pb-Au case with
two different initial temperatures. The background  at low energies consists mainly of $\pi^0$,
$\eta$ and $\omega$ Dalitz decays. The data are taken from \cite{ceres2}.} \label{ceres_pb_au_190_210_mixed}
\end{center}
\end{figure}

\begin{figure}
\begin{center}
\epsfig{file=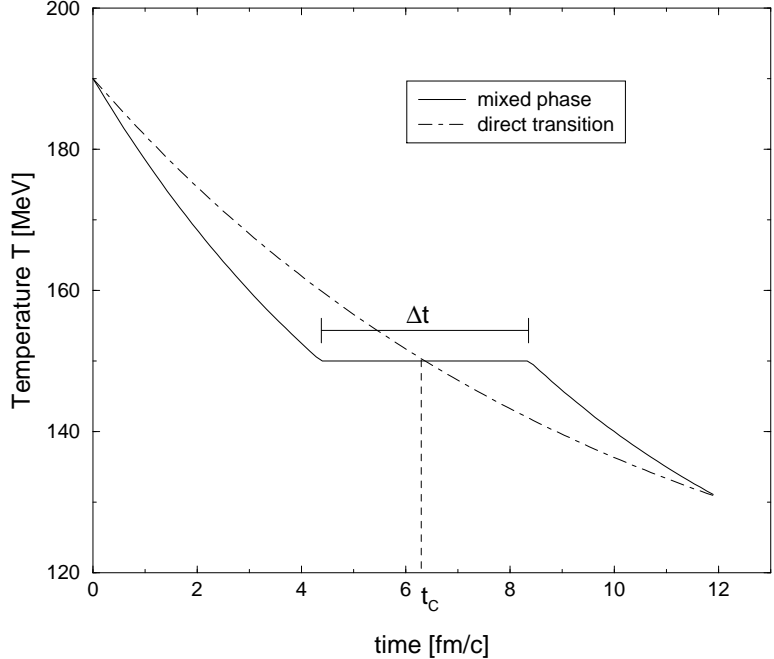,width=9cm, angle = -90} \caption{Fireball temperature as a function of time for a direct transition and a possible mixed scenario.}
\label{temp_profile}
\end{center}
\end{figure}

Whereas the $\rho$ and $\phi$ mesons experience only moderate thermal changes of their spectral
functions, the $\omega$ spectrum changes its shape considerably, mainly due to $\omega \pi \rightarrow \pi\pi$ reactions in the heat bath. We now use eq.(\ref{ImBarPi}) to relate our temperature-dependent vector meson
self-energies to the imaginary part of the electromagnetic current-current correlator which enters
the dilepton rate formula (\ref{dileptonrate}) as long as we stay below $T_C$. Above this critical
temperature, we use expression (\ref{ImPiqqbar}) which yields the dilepton emission rate from a
thermal quark-antiquark system. It is instructive to have a look at the spin-averaged spectral distributions $R(q,T) = 12 \pi \mbox{Im}\bar{\Pi}(q,T) / q^2$ at different temperatures $T$ (presented in fig.\ref{all_spectra} for $q^\mu = (\omega, \vec{q} = 0))$. 
Obviously, the spectral functions on either side of the quark-hadron transition are quite distinct: the $q\bar{q}$ loop spectrum is flat with no significant
structure whereas the hadronic spectra exhibit a clear resonance structure. How these differences
manifest in the dilepton emission spectra is examined in this section where we convolute the
dilepton rate with the space-time evolution of the fireball system to obtain the total rate (\ref{integratedrates}) which can then be compared to the CERES/NA45 data.\\[0.3cm]
\begin{table}[h]
\begin{center}
\begin{tabular}{ll} 
\hline
\noalign{\smallskip}
 & Pb-Au \\
\noalign{\smallskip}\hline\noalign{\smallskip}
$T^i$  & 190 MeV \\
$T^\infty$  & 105 MeV \\
$\tau_1$  & 10 fm \\ 
$\rho^i$ &  2.55 $\rho_0$ \\ 
$\tau_2$  & 6 fm \\ 
$N_B$  & 260 \\ 
$dN/d\eta$  & 220 \\
$t_{f.o.}$ & 10 fm \\ 
$T_{f.o.}$  & 136 MeV \\ 
$\rho_{f.o.}$ & 0.48 $\rho_0$ \\
\noalign{\smallskip}\hline
\end{tabular}  
\caption{Typical set of parameters controlling the space-time evolution of the fireball, adapted for the
Pb-Au collisions at CERN-SPS. Baryon densities are given in units of $\rho_0$ = 0.17 fm$^{-3}$.}\label{fireball_table} 
\end{center}
\end{table}
For a start, we use the parameters shown in table \ref{fireball_table} for eqs.(\ref{Tt}) and
 (\ref{rhot}) which are adapted to reproduce the results of
 microscopic transport calculations \cite{li1}\cite{li2} for the space-time evolution of the S-Au and
Pb-Au collision (we show the Pb-Au parameters as this is the case of primary interest). Let us now
apply our ''mixed'' model scenario (hadronic degrees of freedom below $T_C$, quarks above $T_C$) with $T_C = $ 150
MeV  and initial temperatures of 190 MeV and 210 MeV, respectively.  These two cases are shown
in fig.\ref{ceres_pb_au_190_210_mixed}. For the moment, we employ a direct transition between the two phases at $T_C$ and comment on the implications of a possible mixed phase later. 
We find that we can reproduce the shape of the data quite well. We need a normalization
factor $k$ which can be explained as follows: for the hadronic phase where thermal $\pi\pi$ annihilation
is dominant, there is an increased population of the pion phase space due to a
finite pion chemical potential we did not include \cite{rapp}. However, the striking feature is that
when choosing a higher initial temperature, $T^i = 210 $ MeV, the normalization factor needed to reproduce the data becomes smaller, consistent with the picture that the system now stays for almost the whole expansion time in the
partonic phase where there is no pionic chemical potential. We also checked that our calculations
reproduce the observed transverse momentum distributions: the main part of the enhancement resides in the
$p_T < 500$ MeV region.
\\[0.3cm]
 We can therefore easily account for the dilepton ''excess'' in the intermediate
mass region. At high masses, the scenario with $T^i$ = 190 MeV still underestimates the data a bit.
With a higher initial temperature $T^i = 210$ MeV, the slope becomes flatter and moves up in the
right direction. This can be understood as follows: using a spin-averaged spectral function
as in fig. \ref{all_spectra}, we can rewrite eq.(\ref{dileptonrate}) as
\begin{eqnarray} \frac{dN}{d^4 x d^4q}  & = & \frac{\alpha^2}{12\pi^4} \cdot \frac{R(q^0, \vec{q})}
{e^{\beta q^0} - 1}.\label{spectral}
\end{eqnarray}
From eq.(\ref{ImPiqqbar}) it follows that the thermal spectrum $R$ approaches its constant
perturbative plateau for large momenta and high temperatures or for any momentum at low
temperatures. From this we infer that in the QGP phase, the shape of the dilepton emission rate for
high invariant mass at high $T$ is basically determined by the inverse of the temperature, $\beta$,
as
\begin{equation}
\frac{dN}{d^4 x d^4q}  \sim \mbox{const.} \times e^{-\beta q^0}. \label{ww}
\end{equation}
As the integrated rate (\ref{integratedrates}) averages over all temperatures according to the
fireball parametrization, we see in fig.\ref{ceres_pb_au_190_210_mixed} a straight fall-off at
high invariant masses, with a slope parameter $- \langle 1/T \rangle$, averaged over the
temperature evolution of the fireball. This is an important result: if we start with a different
parametrization of the fireball, say, with a higher initial temperature, {\em the slope changes},
it becomes flatter because we average over $1/T$. Thus an accurate measurement of the dilepton rate
at high invariant mass yields valuable information about the temperature evolution of the fireball,
assuming that equilibration is fast.
\\[0.3cm]
In contrast to its appearance as a sharp resonance at $T=0$, the $\omega$ meson is hardly visible
as a little bump on top of the broad $\rho$ meson spectrum. This was to be expected because the
temperatures of the system in the hadronic phase before freeze-out are still quite high, between 150 and 125 MeV. \\[0.3cm] The
$\phi$ meson can still be identified as a sharp resonance. As the $\phi$ is hardly influenced by
temperature, its height over background ratio may thus act as a chronometer for the critical time
$t_C$ once the fireball parameters are sufficiently well known: at $T^i = 210$ MeV, the $\phi$
meson still sticks out of the QGP 'background', but with a reduced height, confirming that the
system stays now longer in the QGP phase than in the scenario with $T^i = 190$ MeV.\\[0.3cm]
\begin{figure}
\begin{center}
\epsfig{file=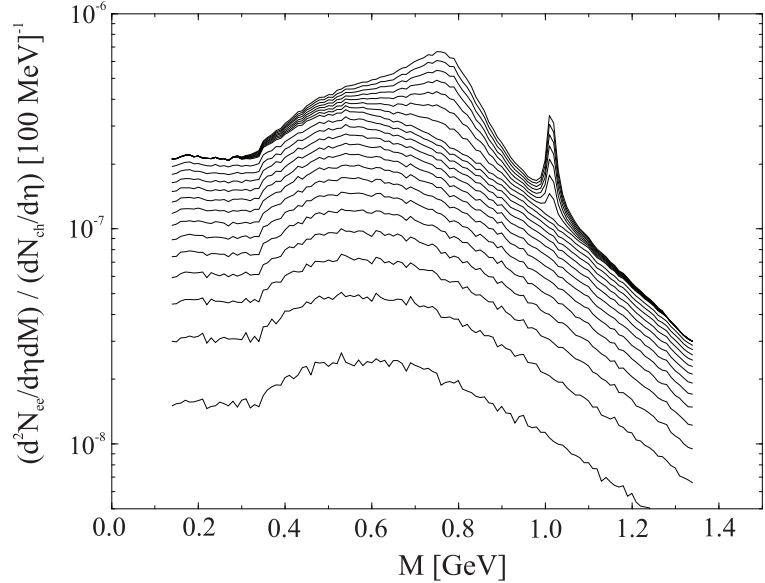,width=10cm} \caption[A time snapshot picture of the Pb-Au
fireball expansion with $T^i$ = 190 MeV]{A time snapshot of the Pb-Au fireball expansion with $T^i$
= 190 MeV (no $k$ factor). Shown are the integrated dilepton rates between the start of the
expansion at $t=0$ and the freeze-out time, $t_f = 10$ fm$/c$ in steps of $\Delta t = $0.5 fm$/c$. The CERES acceptance is
included. The different behaviour of the partonic and the hadronic phases below and above $T_C = $ 150 MeV is clearly visible.}
\label{snapshot_pb_au_190}
\end{center}
\end{figure}
To get a better understanding for the space-time evolution of the fireball, let us look at the time
snapshot picture of the fireball expansion in fig.\ref{snapshot_pb_au_190}. Shown are the
integrated dilepton rates between the start of the expansion and the freeze-out time $t_{f} = 10$
fm$/c$ in steps of $\Delta t = 0.5$ fm$/c$. Although the QGP phase is very hot in the beginning ($t_0 = 0$), its
contribution to the dilepton rate is initially small because it occupies only a small volume. With the
consecutive expansion, the rate per unit volume decreases as the temperature goes down. However,
the increase in volume outweighs this effect partly. The shape of the spectrum stays roughly the
same, but the slope for high $M$ is now, following eq.(\ref{ww}), $- \beta \sqrt{M^2 + p_T^2}$
integrated over the transverse momentum $p_T$. The change in the slope with increasing time, or
equivalently, with decreasing temperature is small yet visible, and the magnitude of the slope is
set by the initial temperature.
\\[0.3cm] At $t_C = 7$ fm$/c$, the phase transition occurs, the system goes into the hadronic phase and the
$\rho$ and $\phi$ meson resonances start to grow out of the QGP 'background'. It is clearly seen
that the shape of the dilepton spectrum at high masses $M$ is only determined by the QGP phase and
that the $\phi$ meson peak height would be larger if the transition occurred earlier or the
freeze-out time were later.\\[0.3cm]
\begin{figure}
\begin{center}
\epsfig{file=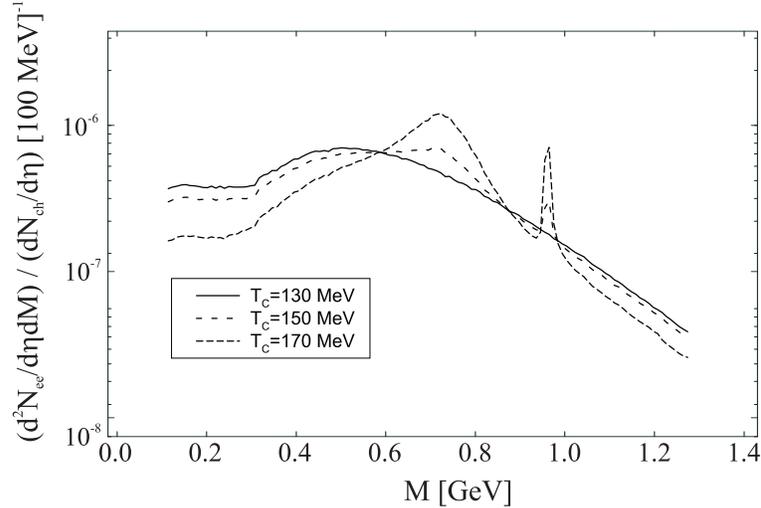,width=10cm} \caption{The resulting Pb-Au dilepton emission spectra
for three different critical temperatures, starting from the same initial temperature $T^i = 210$ MeV. For the sake of lucidity, the low-mass hadronic background has been omitted. No $k$ factor has been applied.} \label{Ti_Tc_210}
\end{center}
\end{figure}
With these observations in mind, we investigate the sensitivity of the spectra with respect to the
parameters $T_C$ and $t_{f}$. In fig.\ref{Ti_Tc_210}, the resulting Pb-Au dilepton emission
spectra are shown for three different critical temperatures $T_C = $130, 150 and 170 MeV. The initial
temperature is 210 MeV. \\[0.3cm] The slopes for high invariant mass $M$ are equal in all three cases of
$T_C$, as expected. For the lowest transition temperature, $T_C$ = 130 MeV, we obtain a pure QGP
spectrum. The higher $T_C$ is chosen,
the more 'hadronic' the spectrum looks and the less dileptons are present in the intermediate mass range
$0.3 - 0.6 $ GeV. At high $T_C$, the broad $\rho$ meson resonance sticks out clearly, so does the narrow $\phi$ meson. From its peak
height it is possible to judge how long the system has stayed in the hadronic phase before
freeze-out. Note that the $\omega$ meson is not visible even for high
$T_C$, because of its increased thermal width. An amusing side note is that all three curves have
four points in common, left and right to the resonances. We also checked that the 
{\em shape} of the dilepton rate is not sensitive to the freeze-out time.
\\[0.3cm]
So far, we have assumed that the transition from the QGP to the hadronic phase is instantaneous, in the sense that the quark phase at $T > T_C$ is separated from the hadron phase at $T < T_C$, and does not affect the time evolution of temperature and volume. Exactly how the hadronization occurs is not known at this stage, but there are indications that it may actually extend over a finite time interval, during which a mixed phase prevails \cite{mixedphase}\cite{kslee}. Lattice data show a discontinuity in the energy density at $T_C$ \cite{latentheat}, so this stage is accompanied by a release of latent heat. Such processes can imply that the temperature stays almost constant during the transition time which may be considerably long, of the order a few fm/$c$ \cite{hwbarz}\cite{eezabrodin}. In order to estimate the consequences of these effects on the dilepton spectra, we have modified the time evolution of the fireball temperature such that there is a plateau of constant temperature $T_C$, as shown in fig.\ref{temp_profile}. The width of the plateau is denoted by $\Delta t$, and the initial and freeze-out temperature are kept at the values listed in table \ref{fireball_table}.
 During the transition, we assume that both phases exist in a homogeneous mixture and their relative occupation of the volume changes with time. Correspondingly, the spectral function entering the dilepton rates in eq.(\ref{spectral}) is modified to 
\begin{equation}
R(t) = \alpha(t) R_{QGP}(T_C) + [1-\alpha(t)]R_{hadr}(T_C),
\end{equation}   
where $\alpha(t_c - \Delta t/2) = 1$, $\alpha(t_c + \Delta t/2) = 0$ and between these limits $\alpha$ drops linearly. We found that even if $\Delta t$ is set to a value as large as 5 fm/$c$, the difference in the dilepton rates is at most 40\%. For $\Delta t = 1$ fm/$c$, the rates are virtually indistinguishable. Within the error bars of the data, we conclude that the existence of a mixed phase is neither confirmed nor excluded at the resolution currently available. 
\section{Conclusions}
Within our simple fireball model, we can indeed reproduce the observed CERES dilepton spectra within
a pure thermal model, assuming that there exists a partonic phase above a certain temperature $T_C$. The
transverse momentum dependence is correctly described, too. With a sufficiently high initial
temperature $T^i = $ 210 MeV, the shape of the spectrum is perfectly well described.
At such high initial temperatures, the system stays in the quark phase for much of its expansion period, and the dilepton radiation originates primarily from quark-antiquark annihilation, resulting in a flat spectrum. Hadronization in the final stage then leads to the appearance of resonance structures. We note that the thermal life time of the $\omega$ meson may decrease such that the $\omega$ actually decays inside the fireball. The detailed shape of the spectra is sensitive to certain combinations of freeze-out, initial and critical temperature.\\[0.3cm] 
An improved experimental mass resolution in the region $M > 1 $ GeV may yield important
information about the initial temperature of the partonic phase. The role of the $\phi$ meson as
both a thermometer and a chronometer for the fireball expansion has been emphasized. Finally we investigated a possible mixed phase scenario and found that it would not leave any distinct traces in the shape of the dilepton rates.
\section{Acknowledgements}
This work has been supported in part by BMBF and GSI.


\begin{thebibliography}{99}

\bibitem{lattice1} F. Karsch, {\em Proceedings of the International Workshop XXV on Gross
Properties of Nuclei and Nuclear Excitations}, Hirschegg, Austria, January 13-18, 1997.
\bibitem{boyd} G. Boyd, S. Gupta, F. Karsch, E. Laermann, B. Peterson and K. Redlich, Phys. Lett. \textbf{B349}, (1995) 170.
\bibitem{lattice2} C. Bernard {\em et al.}, Phys. Rev. \textbf{D55}, (1997) 6861.
\bibitem{recent1} C. Bernard {\em et al.}, Phys. Rev. \textbf{D45}, (1992) 3854.
\bibitem{recent2} Y. Iwasaki {\em et al.}, Z. Phys. \textbf{C71}, (1996) 343.
\bibitem{karsch99} F. Karsch, Nucl. Phys. Proc. Suppl. \textbf{83-84}, (2000) 14.

\bibitem{latentheat} B. Beinlich, F. Karsch and A. Peikert, Phys.Lett. \textbf{B390}, (1997), 268.

\bibitem{ceres1} G. Agakichiev {\em et al.}, CERES collaboration, Phys. Rev. Lett. \textbf{75},
(1995) 1272.
\bibitem{ceres2} G. Agakichiev {\em et al.}, CERES collaboration, Phys. Lett. \textbf{B422},
(1998) 405.
\bibitem{ceres3} B. Lenkeit {\em et al.}, Nucl. Phys. \textbf{A654}, (1999) 627c.

\bibitem{qm99} R. Stock, Nucl.Phys. \textbf{A661}, (1999) 282. 

\bibitem{equ1} K. Geiger, Phys. Reports \textbf{258}, (1995) 237.
\bibitem{equ2} P. Braun-Munzinger, I. Heppe, J. Stachel, Phys. Lett. \textbf{B465}, (1999) 15.
\bibitem{equ3} D. K. Srivastava, {\tt nucl-th/9903066}, subm. to Eur.Phys.J. C

\bibitem{rapp2} R. Rapp, G. Chanfray and J. Wambach, Nucl. Phys. \textbf{A617}, (1997) 472.
\bibitem{li1} Q.G. Li, C.M. Ko and G.E. Brown, Phys. Rev. Lett. \textbf{75}, (1995) 4007; Nucl. Phys.
\textbf{A606}, (1996) 568.
\bibitem{li2} Q.G. Li, C.M. Ko, G.E. Brown and H. Sorge, Nucl. Phys. \textbf{A611}, (1996) 539.  
\bibitem{LeBellac} M. Le Bellac, \textit{Thermal Field Theory} (Cambridge University Press,
1996).
\bibitem{wweise1} F. Klingl, N. Kaiser and W. Weise, Z. Phys. \textbf{A356}, (1996) 193; Nucl. Phys. \textbf{A624}, (1997) 527.
\bibitem{ioffe1} M. Dey, V.L. Eletsky and B.L. Ioffe, Phys. Lett. {\bf B252}, (1990) 620.
\bibitem{ioffe2} V.L. Eletsky, Phys. Lett. \textbf{B245}, (1990) 229.
\bibitem{GSW} M. Gell-Mann, D. Sharp and W.E. Wagner, Phys. Rev. Lett. \textbf{8},
(1952) 261.
\bibitem{ras1} R. A. Schneider and W. Weise, preprint TUM/T39-00-13 (2000).
\bibitem{rapp} R. Rapp and J. Wambach, Adv. Nucl. Phys.(in press), {\tt hep-ph/9909229}, and
references therein; W. Cassing and E.L. Bratkovskaya, Phys. Reports \textbf{308}, (1999) 65. 
\bibitem{mixedphase} K. Rajagopal, Nucl. Phys. \textbf{A661}, (1999) 150.
\bibitem{kslee} K.S. Lee, M.J. Rhoades-Brown and U. Heinz, Phys. Rev. \textbf{C37}, (1988) 1452.
\bibitem{hwbarz} H.W. Barz, B.L. Friman, J. Knoll and H. Schulz, Nucl. Phys. \textbf{A484}, (1988) 661.
\bibitem{eezabrodin} E.E. Zabrodin, L.V. Bravina, L.P. Csernai, H. Stocker and W. Greiner, Phys. Lett. \textbf{B423}, (1998) 373.

\end{thebibliography}
\end{document}